\documentclass[12pt,a4paper]{article}
 \usepackage{epsf}
\begin{document}
\title{The dynamical models and the $Z \rightarrow b \bar{b}$ asymmetry}

\author{Chongxing Yue$^{(a,b)}$, Yuanben Dai$^{(b,c)}$, Hong Li$^{b}$     \\
 {\small a: CCAST (World Laboratory) P.O. BOX 8730. B.J. 100080 P.R. China}\\
 {\small b: College of Physics and Information Engineering,}\\
\small{Henan Normal University, Xinxiang  453002. P.R.China}\\
{\small c: Institute of Theoretical physics, Academia Sinica, P.
O. Box 2735, B.J.100080 P.R. China}
\thanks{This work is supported by the National Natural Science
Foundation of China(I9905004), the Excellent Youth Foundation of
Henan Scientific Committee(9911), and Foundation of Henan
Educational Committee.}
\thanks{E-mail:cxyue@public.xxptt.ha.cn} }
\date{\today}
\maketitle
\begin{abstract}
\hspace{5mm}Motivated by the $3.2\sigma$($1.4\sigma$) deviations
between the recent experimental value for $A_{FB}^{b}(R_{b})$ and
the standard model(SM) prediction, we examine the effect of new
physics(NP) on the $Zb \bar{b}$ couplings $g_L^b$ and $g_R^b$.
First we focus our attention on the dynamical models. Then, using
effective lagrangean techniques, we discuss the corrections of NP
to $g_L^b$ and $g_R^b$. We find some kinds of NP might explain the
recently experimental data about $R_b$ and $A_{FB}^b$. However,
the free parameters of these kinds of NP must be severely
constrained.
\end {abstract}

\vspace{1.0cm} \noindent
 {\bf PACS number(s)}: 12.60Cn,12.60.Nz,13.38.Dg

\newpage
\section{Introduction}
   Most of the experiments are consistent with the predictions of the
standard model(SM) with sufficient accuracy. Almost all of the
experimental data are quite well explained in the context of the
SM. However, in the most recent analysis of the precision
electroweak data\cite{z1}, the $Z \rightarrow b\bar{b}$ forward
-backward asymmetry $A_{FB}^b=0.0990(17)$ is $3.2\sigma$ from the
SM fit value, but there is just a hint of a disagreement, at the
$1.4\sigma$ level, for the $Z\rightarrow b\bar{b}$ branching
ratio $R_b$ $[R_b=0.21664(68)]$. The result presented by the
recent experimental data are very puzzling. Certainly the result
could be a statistical fluctuation or from unknown systematic
error, but Ref[2] has told us that this seems to be due to new
physics(NP) beyond SM. In this note we shall assume that this is
  a signal of new  physics(NP).

   The effective $Zb\bar{b}$ vertex can be parameterized in terms of
 two form factors, the left-handed coupling $g_L^b$ and the right -
 handed coupling $g_R^b$:
 \begin{equation}
 \frac{e}{S_w C_w}[g_L^b\overline{b_L}\gamma^\mu b_L+g_R^b\overline{b_R}
 \gamma^\mu b_R]Z_\mu,
\end{equation}
where $S_w=\sin\theta_w$, $\theta_w$ is the Winberg angle. In order to determine
the two form factors separately, we need two
independent measurement of the $Zb\bar{b}$ vertex. One is provided
by $R_b$ and the other by the forward-backward asymmetry
 $A_{FB}^b$. The deviations of the experimental data about $R_b$ and
 $A_{FB}^b$ from the SM predictions may be induced by the corrections of
 NP to the two from factors $g_L^b$ and $g_R^b$.

In this paper, we shall explore whether the corrections of NP to
the two form factors can bring the theoretical predictions
 closer to the experimental results. We find that some kinds of NP can not
 fit both $R_b$ and $A_{FB}^b$ within the  $1\sigma$ bounds of the recent
  experimental data at the same time. Some other kinds
 of NP might explain the recently experimental data about $R_b$ and
 $A_{FB}^b$. However, the free parameters of these kinds of NP must be
 severely constrained.

\section{Constraints of the experimental data on $g_L^b$ and $g_R^b$ }

   The two form factors $g_L^b$ and $g_R^b$ can be written as :
\begin{equation}
g_L^b=-\frac{1}{2}+\frac{1}{3}S_w^2 +\delta g_L^b, \hspace{5mm}
g_R^b=\frac{1}{3}S_w^2+\delta g_R^b.
\end{equation}
Where $\delta g_{L(R)}^b$ contain the SM and the NP contributions
at one loop order:
\begin{equation}
g_L^b=-\frac{1}{2}+\frac{1}{3}S_w^2+\delta g_L^{b,SM}+\delta
g_L^{b,N} =-0.4208+\delta g_L^{b,N},
\end{equation}
\begin{equation}
 g_R^b=\frac{1}{3}S_w^2+\delta g_R^{b,SM}+\delta g_R^{b,N}=0.0774+
 \delta g_R^{b,N}.
\end{equation}
Where the SM values are for $m_t=174GeV$ and
$M_H=100GeV$\cite{z3}. In principle, the corrections of NP to the
$Zb\bar{b}$ vertex may give arise to one additional form factor,
proportional to $\sigma^{\mu \nu}q^{\nu}$. This magnetic
moment-type form factor arises at one-loop and should be
considered as well. However, its contributions to $R_b$ and
$A^b_{FB}$ are very small. Thus, we have ignored it.

  The corrections of NP to $R_b$ and $A_{FB}^b$ can
be expressed in terms of the form factors $\delta g_L^{b,N}$  and
$\delta g_R^{b,N}$:
\begin{equation}
\frac{\delta R_b}{R_b^{SM}}=2(1-R_b^{SM})\frac{g_L^{b,SM}\delta
g_L^{b,N}+g_R^{b,SM}\delta
g_R^{b,N}}{(g_L^{b,SM})^2+(g_R^{b,SM})^2},
\end{equation}
 \begin{equation}
\frac{\delta
A_{FB}^b}{A_{FB}^{b,SM}}=\frac{4(g_L^{b,SM})^2(g_R^{b,SM})^2}
{(g_L^{b,SM})^4-(g_R^{b,SM})^4}(\frac{\delta
g_L^{b,N}}{g_L^{b,SM}} -\frac{\delta g_R^{b,N}}{g_R^{b,SM}}).
\end{equation}

  The 1$\sigma$ contours of $R_b$ and $A_{FB}^b$ are plotted in Fig.1.
 Since $\frac{\delta R_b}{R_b^{SM}}$ is
more than one order of magnitude smaller than $\frac{\delta
A_{FB}^b}{A_{FB}^{b,SM}}$, the expression $g_L^{b,SM}\delta
g_L^{b,N}+g_R^{b,SM}\delta g_R^{b,N}$ is severely constrained.
From this and that $g_L^{b,SM}$ is about $5.5$ times larger than
$g_R^{b,SM}$ we can see that $\frac{\delta
A_{FB}^b}{A_{FB}^{b,SM}}$ is dominated by the $\frac{\delta
g_R^{b,N}}{g_R^{b,SM}}$ term. Thus, we find that the constraints
of the 1$\sigma$ bounds of $R_b$ and $A_{FB}^b$ on $\delta
g_L^{b,N}/g_L^{b,SM}$ and $\delta g_R^{b,N}/g_R^{b,SM}$ can be
written as:
\begin{equation}
  0.0002\leq\frac{\delta g_L^{b,N}}{g_L^{b,SM}}+(\frac{g_R^{b,SM}}
 {g_L^{b,SM}})^2  \frac{\delta g_R^{b,N}}{g_R^{b,SM}}\leq 0.0041,
 \hspace{5mm}
 0.225 \leq \frac{\delta g_R^{b,N}}{g_R^{b,SM}}\leq 0.465.
\end{equation}
Thus, if the deviations from the SM values about $R_b$ and
$A_{FB}^b$ persist, the corrections to $g_L^b$ and $g_R^b$ from
any kind of NP, which can fit both $R_b$ and $A_{FB}^b$ within the
1$\sigma$ bounds of the experimental at the same time, must
satisfy Eq.(7). This is a very strong constraint.

In the following, we will explore whether the contributions of NP
to the $Zb \bar{b}$ couplings $g_L^{b,SM}$, $g_R^{b,SM}$ can bring
the SM predictions close to the experimental results and see
whether there is any kind of NP satisfying Eq.(7). First we will
mainly focus our attention on the class of models, in which the
electroweak symmetry breaking (EWSB) and the large top mass is
dynamically generated by the new strong interactions. In this
paper, we will call this class of models as the dynamical models.
Then, using effective lagrangean techniques, we discuss the
corrections of NP to the form factors $g_L^b$ and $g_R^b$. This is
a model independent analysis.

\section{The dynamical models}

   To completely avoid the problems arising from the elementary Higgs
field, various kinds of dynamical EWSB mechanisms have been
proposed, and among which topcolor-assisted technicolor (TC2)
theory\cite{z4} is an attractive idea. TC2 theory generally
predicts the existence of two kinds of new gauge bosons:(a) the
extended technicolor (ETC) gauge bosons, (b) the topcolor gauge
bosons including the color-octet colorons $B_{\mu}^{A}$ and an
extra $U(1)_{Y}$ gauge boson $Z^{\prime}$. Furthermore, this kind
of models predict a number of pseudo Goldstone bosons ($PGB^\prime
s$), including the technipions in the technicolor sector and the
top-pions in the topcolor sector. All these new particles can give
corrections to the form factors $g_L^b$ and $g_R^b$.

  The main ETC corrections to $g_L^b$ and $g_R^b$ are from the ETC
gauge boson contributions. It has been shown in Ref.[5] that the
negative diagonal ETC gauge boson contribution to $g_L^b$ is
larger than that of  the positive sideways gauge boson. There
are $\delta g_L^{b,E}<0$ and $\delta g_R^{b,E}\simeq 0$. The gauge
bosons $B_\mu^A$ and $Z^\prime$ also have contributions to $g_L^b$
and $g_R^b$ and there are $\delta g_L^{b,B}/g_L^{b,SM}=\delta
g_R^{b,B}/g_R^{b,SM}>0$, $\delta g_L^{b,Z^\prime}/g_L^{b,SM}>0$
and $\delta g_R^{b,Z^\prime}/g_R^{b,SM}>0$\cite{z6}. Due to the
strong coupling between the top-pion and the third generation quarks
, the top-pions can give rise to a large positive correction to
$g_L^{b,SM}$, i.e. $\delta g_L^{b,\pi_t}/g_L^{b,SM}<0$, $\delta
g_R^{b,\pi_t}\simeq0$\cite{z7}. In Ref.\cite{z8} it is found that,
by combining all of these corrections to $g_L^b$ and $g_R^b$, it is
possible that TC2 models can make the theoretical predictions
consistent with the experimental value of $R_b$. For $A_{FB}^b$,
we have:
\begin{equation}
\frac{\delta
A_{FB}^b}{A_{FB}^{b,SM}}=-\frac{4(g_L^{b,SM})^2(g_R^{b,SM})^2}
{(g_L^{b,SM})^4-(g_R^{b,SM})^4} [\frac{\delta
g_R^{b,B}}{g_R^{b,SM}}
 +\frac{\delta g_R^{b,Z^\prime}}{g_R^{b,SM}}]
\end{equation}
with
\begin{equation}
\frac{\delta g_R^{b,B}}{g_R^{b,SM}}=\frac{k_3}{6\pi}C_2(R)
[\frac{m_Z^2}{M_B^2}\ln \frac{M_B^2}{m_Z^2}],\hspace{5mm}
\frac{\delta
g_R^{b,Z^\prime}}{g_R^{b,SM}}=\frac{k_1}{6\pi}(y_{R}^{b})
[\frac{m_Z^2}{M_{Z^\prime}^2}\ln\frac{M_{Z^\prime}^2}{m_Z^2}].
\end{equation}
We have neglected the  small $\delta g_L^b$ term in above
equations. $k_3$ and $k_1$ are the coloron and the $Z^\prime $
coupling constants, respectively, $M_B$ and $M_{Z^\prime}$ are,
respectively, the mass of $B_{\mu}^A$ and $Z^\prime$,
$C_2(R)=\frac{4}{3}$ and $Y_R^b=-\frac{2}{3}$\cite{z4}. If we take
$k_3=2$, $k_1=1$\cite{z9}, and $M_B=M_{Z^\prime}=500GeV$, we have
$\delta A_{FB}^b\simeq2.6\times 10^{-4}$, which is too small to
explain the $A_{FB}^b$ experimental deviations from the SM
prediction value. Thus, TC2 models can not fit both $R_b$ and
$A_{FB}^b$ within the 1$\sigma$ bounds of the recent experimental
data. If the deviation persist, TC2 models may be ruled out.

  Several years ago, to generate a large top mass, a dynamical model
was proposed in Ref.[10] by B. Holdom. The model contains a fourth
family, and members of the third and fourth families are composed
of two "families" of fermions $ f $ and $\underline{f}$. The model
also predicts the existence of the extra massive gauge boson
$\chi$, which is a singlet under unbroken gauge symmetries and
with a mass in the few hundred GeV to one TeV range. The gauge
boson $\chi$ does not couple to the first and second families. It
couples with a vector charge of $g_\chi$ to all members of the $ f
$ family and with a vector charge of -$g_\chi$ to all members of
the $\underline{f}$ family. The mechanism producing a large top
mass requires that the fourth family quark mass eigenstates
$t^\prime$ and $b^\prime$ correspond to Dirac spinors of the form
[$\underline{f}_L,f_R$], which are nearly degenerate. The t and b
quarks correspond to [ $f_L$, $\underline{f}_R$], which implies
that the gauge boson $\chi$ couples with the same axial coupling
to the t and b quarks. The main effects of $\chi$ on the form
factors $g_L^b$ and $g_R^b$ come from its mixing with the
electroweak gauge boson $Z$, which can be written as [11]:
\begin{equation}
\delta g_L^{b,\chi}=-\delta g_R^{b,\chi}=-\frac{e}{8S_wC_w}
(\frac{m_t}{m_{q^\prime}})^2,
\end{equation}
where $q^\prime=(t^\prime, b^\prime)$. If the dynamical $t^\prime$
and $b^\prime$ masses make the main contributions to the $W$ and
$Z$ masses and the associated decay constant is $F\simeq 145GeV$,
then [11]:
\begin{equation}
 m_{q^\prime}\approx \sqrt{3}F\frac{m_\rho}{2f_\pi}\approx 1TeV .
\end{equation}
Using Eq.(10) and (11), we can easily obtain $\delta g_R^{b,
\chi}/g_R^{b, SM}\simeq3.7\%$, which is too small to explain the
recent experimental data. The reason of generating too small
corrections to $g_R^b$  is that EWSB is mainly induced by the
dynamical $t^\prime$ and $b^\prime$ masses. If we change this
assurance, this problem may be solved. In fact, EWSB may be
induced by two or more kinds of new strong interactions at the
same time or induced by the elementary scalar field. If we assume
that EWSB is driven by the dynamical $t^\prime$ and $b^\prime$
masses and other strong interactions or a Higgs sector, we have:
\begin{equation}
 3(xF)^2+\nu^2={\nu_w}^2,
\end{equation}
with $x$ is a free parameter, $\nu_w \approx 246 GeV $ is the
electroweak scale and $\nu$ represents the contributions of other
strong interactions or a Higgs sector to EWSB. Then, the
corrections of gauge boson $ \chi$ to $g_L^b$  and $g_R^b$ can be
approximately written as:
\begin{equation}
\delta g_L^{b,\chi} \approx -\delta g_R^{b,\chi} \approx
-\frac{e}{8S_w C_w x^2}(\frac{m_t}{1TeV})^2.
\end{equation}

   Using the expression of $ \delta g_R^{b,N}/g_R^{b,SM}$ in Eq.(7),
we can constraint the free parameter $x$, which is in the range of
$0.28 - 0.41$. This means that, to explain the recent experimental
data of  $A_{FB}^b$, the dynamical $t^\prime$ and $b^\prime$
masses must make only small contributions to EWSB and
  the associated decay constant is $F_x\approx 40-60 GeV$.

   The equation (13) gives  too large correction to $\delta g_L^{b}$
as compared to the constraint(7). But, if the scenario described
 above is indeed correct, it can predict the existence of new
scalars with the decay constant $ F_x\approx 40-60 GeV$. Similar
to the top-pions, these new scalars have large Yukawa couplings to
the third family quarks. Thus, these new scalars may have large
positive contribution to $g_L^b$, which can partly cancel the
large negative contributions of the extra gauge boson $ \chi $ to
$ g_L^b$. So this new model might fit both $R_b$ and
 $A_{FB}^b $ within the 1 $\sigma $ bounds of the experimental data
  at the same time.

Other type of new models can also explain the recently
experimental data. For example, the new model proposed by D. Chang
et al\cite{z12} is this case. This new model\cite{z12} contains an
exotic fourth family of quarks and leptons, which is free of
anomalies, together with a heavy Higgs scalar triplet which
supplies the neutrinos with Majorana masses. It has been shown
\cite{z13} that if the top mass $m_t$ is actually larger than
about 230 GeV, and the SM $b_R$ mixes with the exotic quark $Q_1$
of charge -$\frac{1}{3}$  of the doublet $(Q_1,Q_4)_R $, where
$Q_4$ has charge -$\frac{4}{3}$, then this model can account for
all the 1999 precision electroweak data . From Eq.(9) of Ref.[14],
we can see that this  new model can fit both $A_{FB}^b$ and $R_b$
within the 1$\sigma$ bounds of the recent experimental data for
$0.035\leq (\sin{\theta_b})^2\leq 0.072$,  where $\theta_b$ is the
mixing angle of the SM $b_R$ and the exotic quark $Q_1$. The
observed "top quark" phenomenon at the Fermilab are assumed to be
due to $Q_4$.

 \section{Model independent analysis with dimension-six operators}

  In the last several years, many authors \cite{z14,z15} have studied
 the effects of the dimension-six CP conserving $SU(3)_C \times
 SU(2)_L \times U(1)_Y$ invariant operators on the observables
  $R_b$ and $A_{FB}^b$ by using effective Lagrangean techniques. In this
  section, we will use this method to model independent analysis of the
  corrections of NP to the observables $R_b$ and $A_{FB}^b$ and
  compare them  with the recent experimental data.

If we assume that EWSB is dynamical driven by new strong
interactions. This kind of NP may predict the existence of the
operators $O_{qB}$, $ O_{bB} $, in the notation of Ref.\cite{z16}.
These operators arise from the extra $U(1)_Y$ gauge boson B, which
may have significantly contributions to $g_L^b$ and $g_R^b$. The
corrections of the operators $O_{qB}$ and $O_{bB}$ to $g_L^b$,
$g_R^b$ can be explicitly written as:
\begin{equation}
\delta g_L^{b,B}=\frac{2S_w^2
C_w}{e}\frac{C_{qB}}{\Lambda^2}k^2,\hspace{5mm} \delta
g_R^{b,B}=\frac{2 S_w^2 C_w}{e}\frac{C_{bB}}{\Lambda^2}k^2,
\end{equation}
where $\Lambda$ is the NP scale. $k=p_b+p_{\bar{b}}$ is the
momentum of the electroweak gauge boson $Z$ and $C_{ij}$ are
coupling coefficients which represent the coupling strengths of
the operators $O_{ij}$. From Eq.(14), we can see that the
experimental measurement values of $R_b$ and $A_{FB}^b$ can give
severe constrain on the free parameters of NP. If we demand that
 NP can fit both $R_b$ and $A_{FB}^b$ with 1$\sigma $ bounds of
 the recent experimental data at the same time, the value of
 the coupling coefficients $C_{ij}$ can be obtained by using
 Eq.(7). Explicitly
\begin{equation}
0.136\leq C_{qB}\leq 0.604,\hspace{5mm} 1.61\leq C_{bB}\leq 3.32.
\end{equation}

In above estimation, we have taken $\Lambda \approx 1TeV$, Thus,
as long as Eq.(15) is satisfied, it is possible that this kind of
NP models could explain the deviations of the $R_b$ and $A_{FB}^b$
experimental values from the SM predictions.

If we assume that EWSB is driven by elementary scalar fields, the
operators $O_{\phi q}^{(1)}$, $ O_{\phi q}^{(3)} $ and $O_{\phi
b}$ might exist in this kind of NP. Certainly, the operators $O_{b
w \phi}$ and $O_{D b}$ might also exist. However, the
contributions of these operators to $R_b$ and $A_{FB}^b$ are
proportional to $m_b$ and hence are negligible. The corrections of
the operators $O_{\phi q}^{(1)}$, $O_{\phi  q}^{(3)}$ and $O_{\phi
b}$ to the form factors $g_L^b$ and $g_R^b$ can be written as:
\begin{equation}
\delta g_L^{b,\phi}=-\frac{2S_w C_w}{e}\frac{\nu_w m_Z}{\Lambda^2}
(C_{\phi q}^{(1)}+C_{\phi q}^{(3)}),\hspace{5mm} \delta
g_R^{b,\phi}=\frac{2S_wC_w}{e} \frac{\nu_w m_Z} {\Lambda^2}C_{\phi
b}.
\end{equation}
 Using Eq.(7) and (16), we can obtain:
\begin{equation}
-0.108 \leq C_{\phi q}^{(1)}+C_{\phi q}^{(3)}\leq -0.024,
\hspace{5mm} 0.287\leq C_{\phi b}\leq 0.593,
\end{equation}
which is required to have the theoretical values of both $R_b$ and
$A_{FB}^b$ to lie within the $1\sigma$ bounds of the recent
experimental data.

   From Eq.(16) and (17), we can see that the contributions of the
operators $O_{\phi q}^{(1)}$, $O_{\phi q}^{(3)}$ and $O_{\phi b}$
to $R_b$ and $A_{FB}^b$ are larger than those of the operators
$O_{q B}$ and $O_{b B}$. The NP models which can predict the
operators $O_{\phi q}^{(1)}$, $O_{\phi q}^{(3)}$ and $O_{\phi b}$
are more severely  constrained by the experimental data. However,
with appropriate parameter values, all of two kinds of NP models
can fit both $R_b$ and $A_{FB}^b$ within the $1\sigma$ bounds of
the experimental data at the same time.

\section{Conclusions }

  The effective $Zb\bar{b}$ vertex can be parameterized in terms of
two form factors $g_L^b$ and $g_R^b$. These two form factors can
be determined by the observables $R_b$ and $A_{FB}^b$. Thus, using
the new  results of $R_b$ and $A_{FB}^b$, we can obtain the
constrains of the experimental data on $\delta g_L^{b,N}$ and
$\delta g_R^{b,N}$. On this basis, we discuss the contributions of
some kinds of NP to $\delta g_L^b$  and $\delta g_R^b$, we find
that some  models, such as TC2 models, can not fit both $R_b$ and
$A_{FB}^b $ within the $1\sigma$ bounds of new experimental data
at the same time. However, some kinds of NP, for example, some
modification of the model proposed in Ref.\cite{z11} might explain
the deviations of the new experimental data about observables
$R_b$ and $A_{FB}^b$ from the SM predictions. Certainly, in the
framework of these kinds of NP, model building must be studied
extensively in the future. Lastly, we use effective Lagrangean
techniques to model independent analysis of the corrections of NP
to the observables $R_b$ and $A_{FB}^b$ and compare them  with the
recent experimental data. We find that, with appropriate parameter
values, some kinds of NP models can fit the experimental data of
the observables $R_b$ and $A_{FB}^b$ at the same time.

\newpage
\vskip 2.0cm
\begin{center}
{\bf Figure captions}
\end{center}
\begin{description}
\item[Fig.1:]The 1$\sigma$ contours for $R_b$ and $A_{FB}^b$  in the $\delta g_L^b-\delta
g_R^b$ plane. The solid line represents that the contribution of
NP makes the SM prediction value of $R_b$ and $A_{FB}^b$ to the centre value of
the experimental data.

\end{description}

\newpage
\begin{figure}[pt]
\begin{center}
\begin{picture}(250,200)(0,0)
\put(-50,20){\epsfxsize120mm\epsfbox{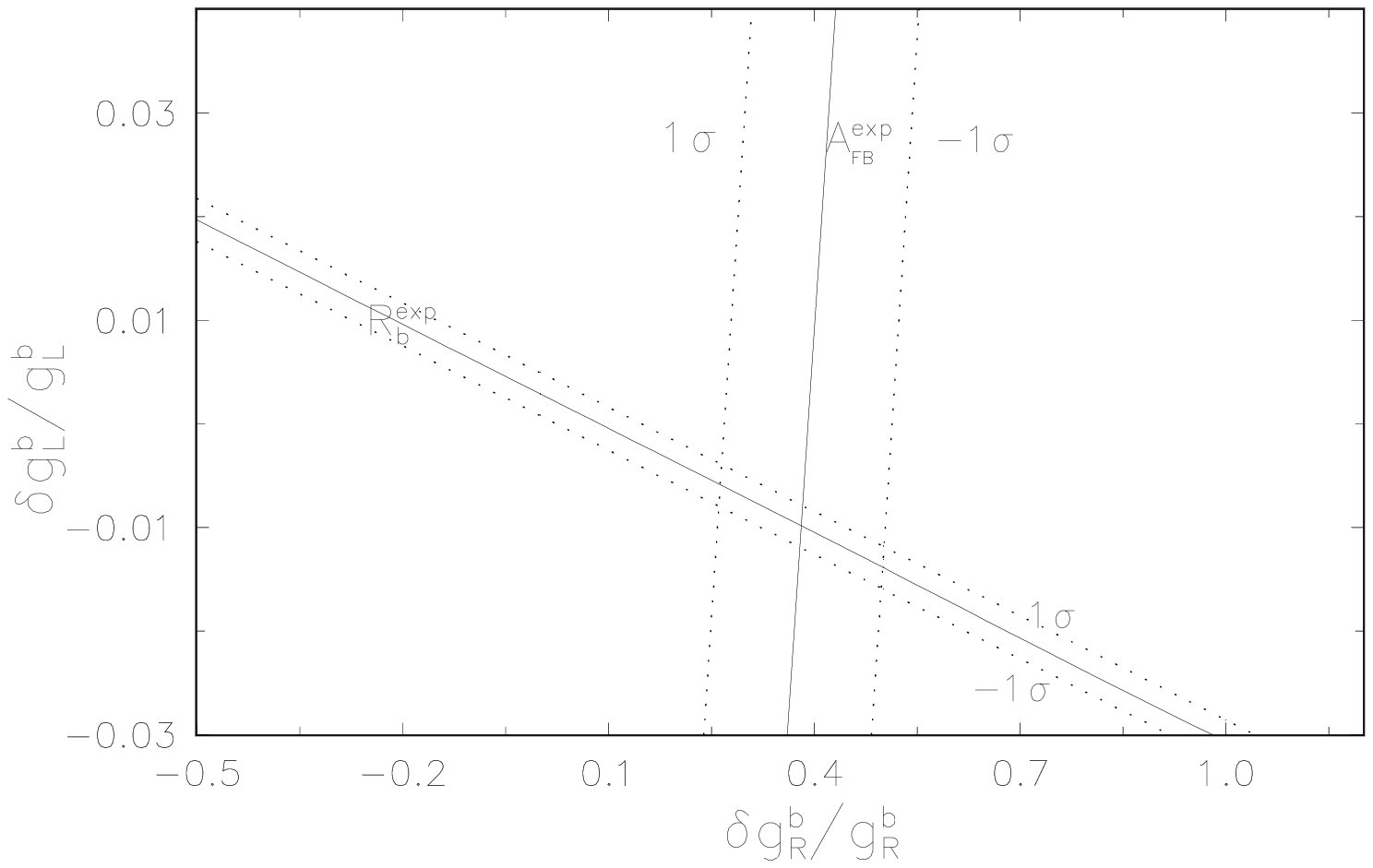}}
 \put(120,-10){Fig.1}
\end{picture}
\end{center}
\end{figure}

\end{document}